\documentclass[a4paper,11pt]{article}
\usepackage{pos}
\usepackage{subcaption}

\title{On the Origin of Ultra-high-energy Cosmic Rays Assuming a Heavy Mass Composition}

\author*[a]{Alena Bakalová}
\author[a]{Ana Laura Müller}
\author[a]{Jakub Vícha}
\affiliation[a]{Institute of Physics of the Czech Academy of Sciences, Prague, Czech Republic}

\emailAdd{bakalova@fzu.cz}

\abstract{Recent studies, supported by updated hadronic interaction models, suggest that the mass composition of ultra-high-energy cosmic rays may be heavier than previously assumed. This has significant implications for source identification, as the deflections of the Galactic magnetic field (GMF) are larger for heavy primaries than for lighter ones at the same energy. In this work, we assume that cosmic rays above 40~EeV consist of iron nuclei only and investigate their possible sources through simulations of cosmic ray propagation, including interactions with ambient photon fields and deflections in the GMF using multiple models. We consider two types of sources as potential origins of these cosmic rays, active galactic nuclei and starburst galaxies. We compare the predicted distributions of arrival directions from sources within 250~Mpc with the measured arrival directions of cosmic rays above 40~EeV. Our results indicate that stronger correlation is found for the active galactic nuclei scenario compared to starburst galaxies. However, we find that within our heavy mass composition model, the GMF leads to significant deflections, making source identification challenging with current knowledge and tools, even at the highest energies.}

\ConferenceLogo{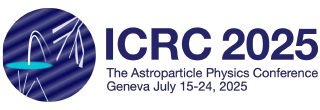}

\FullConference{39th International Cosmic Ray Conference (ICRC2025)\\
 15–24 July 2025\\
Geneva, Switzerland\\}

\begin{document}
\maketitle

\section{Introduction}


While the origin of ultra-high-energy cosmic rays (UHECRs) remains an open question in astroparticle physics, recent measurements offer valuable clues about their possible sources. The dipole anisotropy in arrival directions of UHECRs above 8 EeV suggests their extragalactic origin \cite{AugerDipole2017} and data from the Pierre Auger Observatory show a correlation between arrival directions of UHECRs above 40~EeV and extragalactic objects, with the strongest correlation found for starburst galaxies \cite{CorreleationsAuger2018, AugerADs2022ApJ}. However, in these correlation studies the effect of the Galactic magnetic field is not taken into account. 

During their propagation from sources to the Earth, UHECRs can interact with cosmic microwave background (CMB) and the extragalactic background light (EBL).  These interactions not only lead to energy losses, but in the case of nuclei heavier than protons, can also alter their mass composition. Additionally, such interactions limit the horizon from which the most energetic particles might originate. Cosmic rays are also deflected in magnetic fields, including the Galactic magnetic field (GMF) and extragalactic magnetic field (EGMF). As a result, tracing UHECRs back to their sources remains a major challenge in astroparticle physics.

We follow up on the heavy metal scenario, an extreme scenario that assumes the mass composition of cosmic rays above 40 EeV consists of pure iron nuclei \cite{Vícha_2025}. In this study, we assume the pure iron nuclei above 40 EeV to be emitted from the sources rather than propagated up to the Earth, which allows us to account for propagation effects such as photo-disintegration and energy losses during cosmic-ray propagation in the extragalactic space. 

We analyze the patterns in the arrival directions of cosmic rays above 40~EeV from selected catalogue sources for two scenarios of UHECR origin; active galactic nuclei (AGNs) and starburst galaxies (SBGs), while taking into account their interactions in the extragalactic space and deflections in the GMF. These predicted flux distributions are then compared with the observed arrival directions of cosmic rays above 40~EeV measured by the Pierre Auger Observatory \cite{AugerADs2022ApJ}.

Compared to our previous results \cite{BakalovaUHECR24}, we extend the analysis by including extragalactic propagation and interactions of cosmic rays with CMB and EBL. Furthermore, we improve the modeling of deflections in the GMF by employing high-resolution magnetic lenses for three models of the GMF. Finally, for the selected AGN sources, we utilize the updated \textit{Fermi}-LAT 4LAC catalogue \cite{Fermi4LAC}.

\section{Catalogue sources}
\label{sec:sources}
For the AGN selection, we use the \textit{Fermi}-LAT 4LAC catalogue of active galactic nuclei. Only sources within a distance of 250~Mpc are considered. The distance is calculated using the \texttt{astropy.cosmology} module with a flat $\Lambda$CDM model ($H_0 = 70$~km\,s$^{-1}$\,Mpc$^{-1}$, $\Omega_m = 0.3$). For nearby sources within 40~Mpc, we adopt redshift-independent distances from the NED\footnote{\url{https://ned.ipac.caltech.edu/}} database. Several sources are excluded due to significant discrepancies between the redshift values listed in the 4LAC catalogue and those in NED. Additionally, sources classified as \textit{bcu} (blazar candidates of uncertain type) are removed from the sample. The final selection consists of 55 AGNs. 

In the propagation simulations, all sources are assumed to emit UHECRs with equal intrinsic power. The energy flux in the 100~MeV to 100~GeV range is then used as a proxy for UHECR emission to weight each source’s contribution in the final flux maps. However, in the case of blazars, the observed luminosity is higher compared to their intrinsic luminosity due to relativistic beaming and should be corrected in order to be used as a UHECR proxy \cite{deOliveira_2025}. We correct the observed luminosity of blazars $\mathcal{L^{\rm{obs}}}$ to an intrinsic luminosity $\mathcal{L^{\rm{in}}}$ as
\begin{equation}
   \mathcal{L^{\rm{in}}}= \mathcal{D}^{-2}\mathcal{L^{\rm{obs}}},
\end{equation}
where $\mathcal{D}$ is the Doppler factor. The Doppler factor is taken from \cite{deOliveira_2025, YeDoppler} when available and otherwise estimated using the approximation from \cite{Chen23DopplerAprox}.

For the SBG sample we adopt the selection from \textit{Lunardini et al.}~2019 (Table 1 in \cite{Lunardini2019}) but excluding the Large and Small Magellanic Clouds. The final sample consists of 43 SBGs within $\sim 100$ Mpc. For this scenario, we use the radio flux as a proxy for the UHECRs emission.

\section{Propagation of cosmic rays}

The propagation of cosmic rays is modeled in two stages, treating extragalactic and Galactic propagation separately. Firstly, at the extragalactic stage, energy losses due to interactions with background photon fields and cosmological redshift are simulated. Secondly, the Galactic  propagation models the deflections in the Galactic magnetic field.

\subsection{Simulations of extragalactic propagation}

The extragalactic propagation of cosmic rays is simulated with CRPropa~3 \cite{CRPropa32} using one-dimensional propagation. The simulation accounts for interactions with the CMB and EBL (Gilmore 12 model \cite{Gilmore12}), as well as cosmological redshift.  

All sources are modeled as identical, emitting iron nuclei from 40~EeV up to 300~EeV with a power-law energy spectrum with a broken exponential rigidity cutoff as

\begin{equation}
       \frac{\mathrm{d}N}{\mathrm{d}E}\propto E^{-\gamma}\cdot f_{\rm{cut}},
\end{equation}
where $\gamma$ is the spectral index on the source and $f_{\rm{cut}}$ is the rigidity dependent cutoff function

\begin{eqnarray}
 f_\text{cut} = \begin{cases}
  1 &(E<Ze R_\text{cut})  \\
  e^{\left(1-\frac{E}{Ze R_\mathrm{cut}} \right)} & \left( E>Ze R_\mathrm{cut} \right) \label{equation:2}
 \end{cases},
\end{eqnarray}
where $Z$ is the proton number of the particle and $R_{\rm{cut}}$ is the rigidity cutoff. In order to describe the energy spectrum on the Earth well, the spectral features $\gamma$ and $R_{\rm{cut}}$ on the sources are fitted to match the energy spectrum above 40 EeV measured by the Pierre Auger Observatory \cite{PAOenergyspectrum}, separately for AGNs and SBGs. 

\subsection{Simulations of Galactic propagation}

Particles obtained from all the catalogue sources are merged together and reweighted based on the flux of a given source (see Section~\ref{sec:sources}). To model deflections in the EGMF, the arrival directions at the edge of the Galaxy are sampled from a von Mises–Fisher distribution around the source direction with a blurring angle $\delta_{\mathrm{EGMF}}$. The blurring angle depends on the rigidity $R$ of the arriving particle on the edge of the Galaxy as
\begin{equation}
\label{eq:blurring}
    \delta_{EGMF}(R)=\delta_0\frac{1}{R[\rm{EV}]/40\,\rm{EV}},
\end{equation}
where $\delta_0$ corresponds to the deflection of a proton with energy 40~EeV.

The sampled directions at the edge of the Galaxy are then propagated through the GMF using high-resolution magnetic lenses to obtain their arrival directions at Earth.  We use magnetic lenses for three models of the GMF, the KST24 model \cite{KST24}, UF23 base model \cite{UF23} and JF12 model with Planck tuned parameters \cite{JF12, JF12Planck}, all with the turbulent component from the Planck tuned JF12 model. Each lens is defined using a HEALPix\footnote{\url{http://healpix.sourceforge.net}} binning with nside=64.

\section{Results}

\subsection{Flux maps from selected catalogue sources}

\begin{figure}[!h]
 \centering
 \begin{subfigure}[h]{0.49\textwidth}
     \centering
     \includegraphics[width=\textwidth]{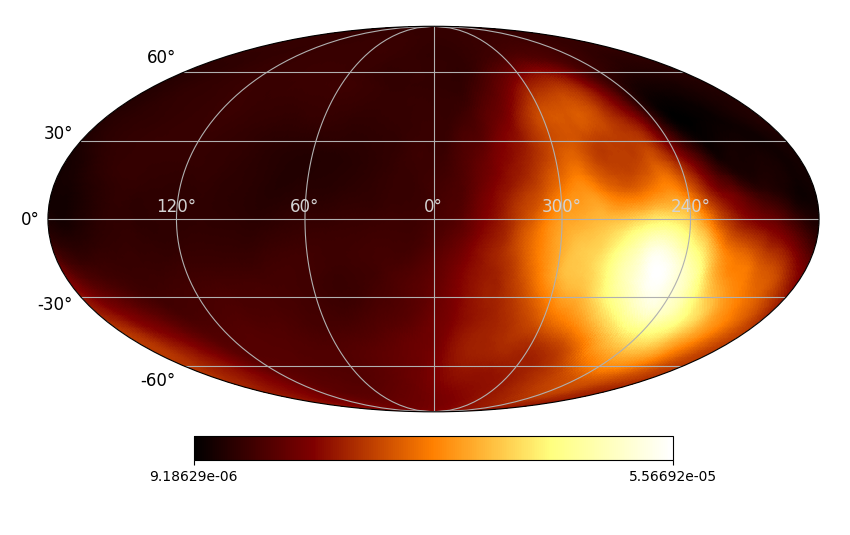}
     \caption{AGNs, KST24}
     \label{..}
 \end{subfigure}
 \hfill
 \begin{subfigure}[h]{0.49\textwidth}
     \centering
     \includegraphics[width=\textwidth]{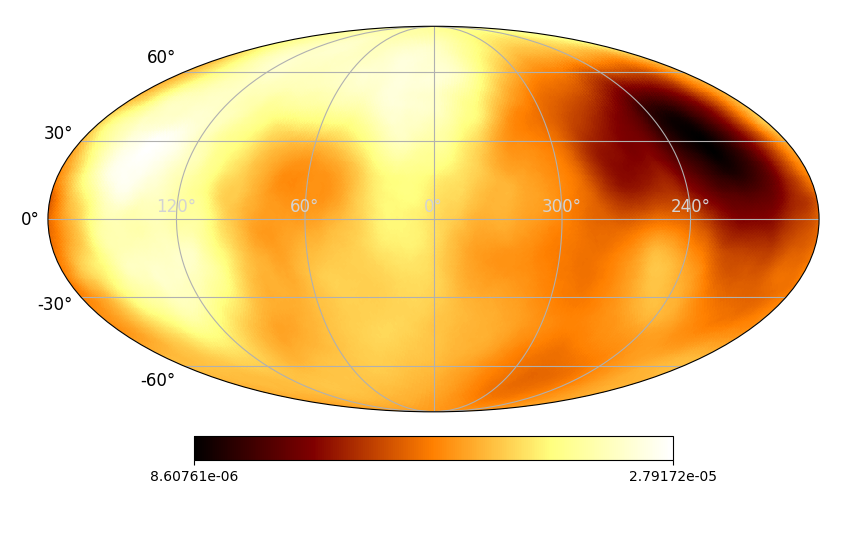}
     \caption{SBGs, KST24}
     \label{..}
 \end{subfigure}
 \hfill
 \begin{subfigure}[h]{0.49\textwidth}
     \centering
     \includegraphics[width=\textwidth]{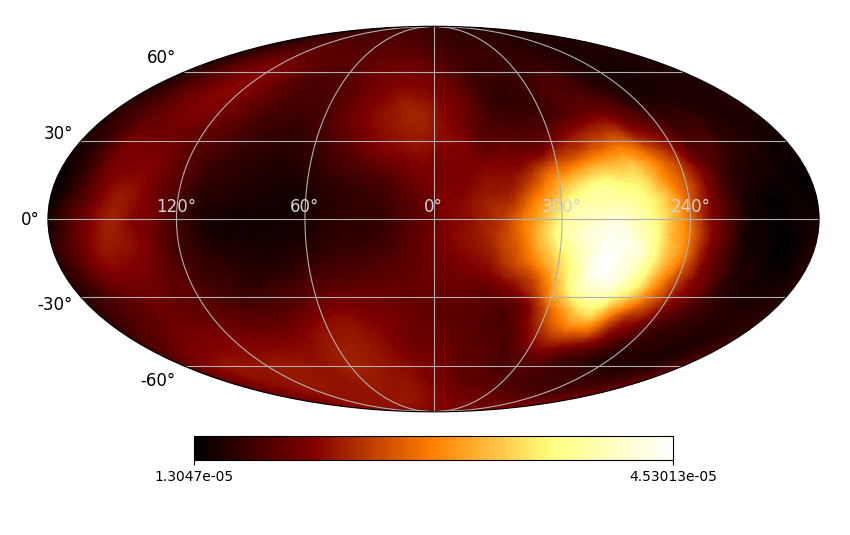}
     \caption{AGNs, UF23 base}
     \label{..}
 \end{subfigure}
 \hfill
 \begin{subfigure}[h]{0.49\textwidth}
     \centering
     \includegraphics[width=\textwidth]{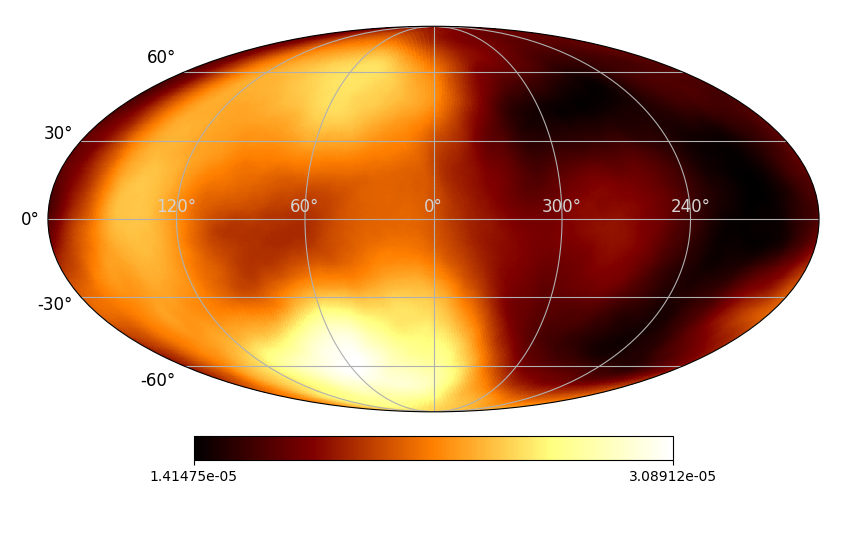}
     \caption{SBGs, UF23 base}
     \label{..}
 \end{subfigure}
 \hfill
 \begin{subfigure}[h]{0.49\textwidth}
     \centering
     \includegraphics[width=\textwidth]{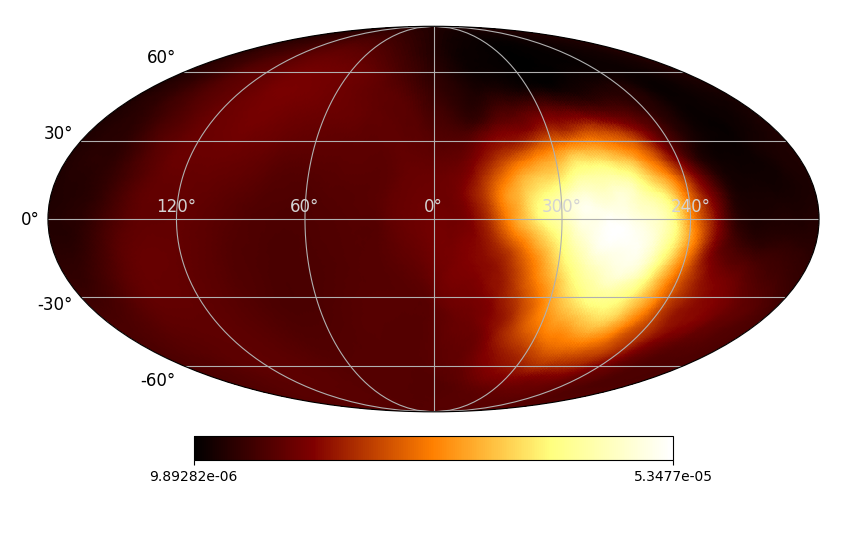}
     \caption{AGNs, JF12Planck}
     \label{..}
 \end{subfigure}
 \hfill
 \begin{subfigure}[h]{0.49\textwidth}
     \centering
     \includegraphics[width=\textwidth]{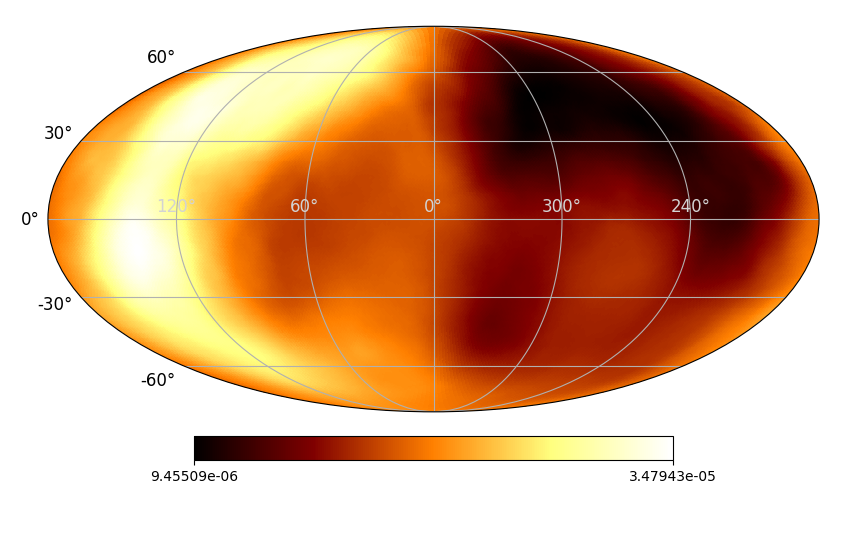}
     \caption{SBGs, JF12Planck}
     \label{..}
 \end{subfigure}
 \caption{Normalized flux maps on Earth of cosmic rays above 40~EeV from selected catalogue sources, AGNs (left) and SBGs (right) assuming pure iron nuclei emitted at the sources with the effects of the GMF, using KST24 field with JF12Planck turbulent field (top),  UF23 base field with JF12Planck turbulent field (middle) and JF12Planck model (bottom). The EGMF is considered with blurring angle $\delta_{EGMF}$ from Equation~\eqref{eq:blurring} with $\delta_{0}=1^{\circ}$ and the flux maps are smoothed with a $15^{\circ}$ top-hat function.}
    \label{fig:FluxMaps}
\end{figure}

Figure~\ref{fig:FluxMaps} shows the predicted flux maps at Earth for the AGN and SBG source scenarios, using a blurring angle $\delta_0=1^{\circ}$ and the three models of the GMF. The flux maps are smoothed by a $15^{\circ}$ top-hat function.

In both scenarios, the flux is dominated by nearby sources. In the case of the AGNs,  the largest contribution to the flux of cosmic rays above 40~EeV comes from Centaurus~A (at a distance of $\sim$3.8~Mpc), while other sources with smaller contributions include M~87, Fornax~A or NGC~1275. In the case of the SBGs scenario, the flux on the Earth is dominated by M~82 (at $\sim 3.6$~Mpc) with additional contributions from NGC~4945, NGC~253 or NGC~1068. While a flux excess is observed near the Centaurus A region in the AGN scenario, the SBG scenario produces only a weak local excess in that area (originating from NGC 4945),  with most of the flux instead concentrated toward positive galactic longitudes, assuming iron nuclei at the sources above 40~EeV.

\subsection{Correlation with arrival directions of cosmic rays}

We compare the flux maps from selected catalogue source for the AGN and SBG scenarios of arrival directions above 40~EeV with the published arrival directions of cosmic rays above 40~EeV \cite{AugerADs2022ApJ} assuming pure iron nuclei at the selected sources. The modeled flux maps are reweighted by the Pierre Auger Observatory exposure of the Surface Detector array \cite{SOMMERS_SDexposure}. We perform a likelihood ratio test of the arrival directions of cosmic rays with isotropy as a null hypothesis $\mathcal{L}_{\rm{iso}}$ against the source model $\mathcal{L}_{\rm{model}}$ separately for the two scenarios. The source models consist of flux from the catalogue sources with relative fraction $f_{\rm{cat}}$, combined with an isotropic background of fraction $(1 - f_{\rm{cat}})$. The test statistic TS, defined as

\begin{equation}
    TS=-2( \ln\mathcal{L}_{\rm{iso}}- \ln\mathcal{L}_{\rm{model}}),
\end{equation}
is minimized using two free parameters, $f_{\rm{cat}}$ and $\delta_{0}$. The parameter space is scanned with steps of $0.2^{\circ}$ in $\delta_0$ and 5\% in $f_{\rm{cat}}$. 

In the case of the AGNs, a positive TS was found for all three models of the GMF,  indicating a mild preference over isotropy. The maximum test statistic $TS_{\rm{max}}$ and the corresponding values of $f_{\rm{cat}}$ and $\delta_0$ are summarized in Table~\ref{tab:fits} for each GMF model. The highest TS is found for the JF12Planck model of the GMF, with $TS_{\rm{max}}=7.62$, a catalogue fraction 15\% and blurring angle  $\delta_{0}=1.0^{\circ}$. It is important to note that this blurring angle corresponds to the deflection of protons with energy 40~EeV. In our scenario, however, the flux is dominated by nuclei close to iron, which experience deflections $Z$ times larger due to their higher charge/lower rigidity.

In the case of the SBGs, the TS is consistently negative for all three models of the GMF. This is primarily due to the concentration of flux in regions of positive Galactic longitude and the lack of significant excess near the Centaurus~A region. In particular, the expected flux from NGC~4945, a prominent SBG located near Centaurus~A, is substantially reduced due to de-magnification effects in the GMF when assuming iron nuclei and the proxy used for its UHECR emission cannot compensate for the suppression caused by magnetic deflections.



\begin{table}[h!]
\centering
\begin{tabular}{|c|c|c|c|c|}
\hline
Scenario & GMF & $TS_{\rm{max}}$ & $\delta_0$ [$^{\circ}$] & $f_{\rm{cat}} [\%]$   \\ \hline
AGNs    & KST24 & 4.32 &$0.4$             & 10  \\ \hline
AGNs    & UF23 base & 7.35 &$0.6$             & 25  \\ \hline
AGNs    & JF12Planck & 7.62 & $1.0$             &  15  \\ \hline
\end{tabular}
\caption{Maximum test statistic $TS_{\rm{max}}$ and corresponding values of the blurring angle $\delta_\mathrm{EGMF}$ and catalogue fraction $f_{\rm{cat}}$ for the AGN scenario and the three models of the GMF.}
\label{tab:fits}
\end{table}

\section{Summary}
We study the arrival directions of cosmic rays above 40 EeV from selected catalogue sources of active galactic nuclei and starburst galaxies. We assume pure iron nuclei at the sources, following up on the Heavy metal scenario~\cite{Vícha_2025}. The simulation of cosmic ray propagation is divided into two stages; an extragalactic propagation of energy losses of cosmic rays on the CMB and EBL and the Galactic propagation taking into account deflections in the GMF. Three models of the GMF are studied, KST24, UF23 base and JF12Planck, all with a turbulent component from the Planck-tuned JF12 model.

We produce flux maps of UHECRs on Earth for both source scenarios. In the AGN case, the flux is dominated by Centaurus~A, with smaller contributions from sources such as M~87, Fornax~A, and NGC~1275. For the SBG scenario, the flux is primarily driven by M~82, while the contribution from NGC~4945 is suppressed due to de-magnification effects in the GMF, despite its proximity and comparable radio flux.

We compare the modeled flux maps with the observed arrival directions of cosmic rays above 40~EeV measured by the Pierre Auger Observatory. In the AGN scenario, a mild preference over isotropy is found for all three GMF models. In contrast, the SBG scenario shows no significant improvement over isotropy in any of the GMF models and the TS is consistently negative over the whole parameter space, when the iron nuclei are assumed at the sources.

\section*{Acknowledgments}
The work was supported by the Czech Academy of Sciences: LQ100102401, Czech Science Foundation: GACR 24-13049S, Ministry of Education, Youth and Sports, Czech Republic: FORTE
CZ.02.01.01/00/22\_008/0004632.   
This research has made use of the NASA/IPAC Extragalactic Database (NED),
which is operated by the Jet Propulsion Laboratory, California Institute of Technology, under contract with the National Aeronautics and Space Administration. Some of the results in this paper have been derived using the healpy and HEALPix packages.

\bibliographystyle{JHEP}
\bibliography{biblio.bib}


\end{document}